\documentclass[aps,prl,twocolumn,superscriptaddress]{revtex4}
\usepackage{graphicx}
\usepackage{color}
\usepackage{amsmath,amssymb,amsfonts}
\usepackage{bm}
\usepackage{gensymb}
\usepackage{tikz}
\usepackage{url}

\newcommand{\nq}{N}
\newcommand{\ns}{N_{\mathrm{\text{SWAP}}}}
\newcommand{\nd}{N_{d}}
\newcommand{\dx}{d_{x}}
\newcommand{\dy}{d_{y}}

\newcommand{\lx}{l_{x}}
\newcommand{\ly}{l_{y}}

\newcommand{\linear}{\mathrm{lin}}
\newcommand{\rect}{\mathrm{rec}}
\newcommand{\twod}{\mathrm{2d}}

\begin{document}


\title{Compilation and scaling strategies for a silicon quantum processor \\ 
with sparse two-dimensional connectivity}




\author{O.~Crawford}
\email[]{ophelia.crawford@riverlane.com}
\affiliation{Riverlane, Cambridge, United Kingdom}
\author{J.~R.~Cruise}
\affiliation{Riverlane, Cambridge, United Kingdom}
\author{N.~Mertig}
\affiliation{Hitachi Cambridge Laboratory, J. J. Thomson Ave., Cambridge, CB3 
0HE, United Kingdom}
\author{M.~F.~Gonzalez-Zalba}
\email[]{mg507@cam.ac.uk}
\thanks{Present Address: Quantum Motion Technologies, Cornwall Road, Harrogate, HG1 2PW, United Kingdom}
\affiliation{Hitachi Cambridge Laboratory, J. J. Thomson Ave., Cambridge, CB3 
0HE, United Kingdom}


\date{\today}

\begin{abstract}

Inspired by the challenge of scaling up existing silicon quantum hardware, we 
investigate compilation strategies for sparsely-connected 2d qubit arrangements 
and propose a spin-qubit architecture with minimal compilation overhead.
Our architecture is based on silicon nanowire split-gate transistors which can form finite 1d 
chains 
of spin-qubits and allow the execution of two-qubit operations such as \textsc{Swap} 
gates among neighbors.
Adding to this, we describe a novel silicon junction which can couple up to four
nanowires into 2d arrangements via spin shuttling and \textsc{Swap} 
operations.
Given these hardware elements, we propose a modular sparse 2d spin-qubit architecture 
with unit cells consisting of diagonally-oriented squares with nanowires along 
the edges and junctions on the corners.
We show that this architecture allows for compilation strategies which 
outperform the best-in-class compilation strategy for 1d chains, not only 
asymptotically, but also down to the minimal structure of a single square.
The proposed architecture exhibits favorable scaling properties which allow for 
balancing the trade-off between compilation overhead and colocation of 
classical 
control electronics within each square by adjusting the length of the nanowires.
An appealing feature of the proposed architecture is its manufacturability
using complementary-metal-oxide-semiconductor (CMOS) fabrication processes.
Finally, we note that our compilation strategies, while being inspired by 
spin-qubits, are equally valid for any other quantum processor with sparse 2d 
connectivity.

\end{abstract}


\maketitle

\section{Introduction}

Qubit connectivity is a primary feature of any quantum computing technology.
It represents the architectural arrangement of the qubits within the quantum 
processor and indicates the number of qubits with which any other qubit can interact.
Highly-connected structures are favorable since two-qubit gates between 
arbitrary qubits require a smaller gate count and hence more complex problems 
can be solved with lower circuit depth.
On the other hand, high qubit connectivity comes at the expense of 
technological 
complexity.
Therefore, scaling while maintaining high-connectivity is a challenge being faced by 
many quantum computing technologies.
All-to-all connectivity has been demonstrated by photonic~\cite{Zhong2020} as 
well as ion-trap~\cite{Linke2017, Pino:2021aa} qubits,
but the distributed nature of these technologies puts serious challenges on 
the path to scaling.
On the other hand, solid-state systems, such as superconducting~\cite{Arute2019} 
and quantum-dot spin qubits~\cite{xue2021a}, can exhibit 2d hardware 
topologies, which could be compactly integrated on a chip.
This allows for colocating classical support electronics and offers an 
alternative path to scaling.

Implementations of 2d grids with nearest neighbor 
connectivity are desirable as this would allow for fault-tolerant quantum 
computing via the surface code~\cite{Fowler2012} without additional gate 
overhead.
However, scaling 2d nearest neighbor grids in solid-state systems poses substantial contact routing 
challenges and complicates the integration of classical support electronics in 
the qubit plane~\cite{Charbon2016, Gonzalez-Zalba2020}.
Quantum-classical integration would require levels of 3d integration unseen to date~\cite{Veldhorst2017}. 
Therefore, solid-state platforms explore scaling with sparse 2d 
connectivity~\cite{Vandersypen2017, Buonacorsi2019, Nation2021} and several 
works have explored optimized methods for running quantum algorithms on 
sparsely-connected hardware graphs~\cite{Kivlichan2018, Guerreschi2018, 
Holmes2020}.

Particularly for the spin qubits, progress has been made in the last 
few years demonstrating high-fidelity gates~\cite{Yoneda2017,Yang2019} and 
readout~\cite{HarveyCollard2018, Urdampilleta2019} beyond the fault-tolerance 
threshold~\cite{xue2021a, noiri2021, mills2021}. Further, the first few qubit processors \cite{Zajac2017, 
Hendrickx2021} and blueprints \cite{Vandersypen2017, Veldhorst2017, Li2018, 
Boter2021} are beginning to emerge.
Next steps will be focused towards scale-up, for which 
the use of industrial complementary metal-oxide-semiconductor (CMOS) processes is expected to play an important 
role~\cite{Veldhorst2017, Gonzalez-Zalba2020}.
In particular, modules containing industry-manufactured bilinear arrays of 
quantum dots (QDs) using split-gate nanowire transistor technology are readily 
manufactured and their qubit properties are widely being tested in 
experiments~\cite{Betz2016, Hutin2019, Ansaloni2020, Chanrion2020}.
These modules should provide a platform to demonstrate a 1d quantum processor 
in 
silicon which should allow for running early noisy intermediate-scale quantum (NISQ) 
algorithms~\cite{Kivlichan2018, Cai2020}, Shor's algorithm~\cite{Fowler2004} or 
demonstrating logical qubits~\cite{Jones2018}.
How best scale this technology to allow 
for optimized operation of quantum algorithms with minimal compilation overhead
is therefore an important question.

Here, we present a concept to utilize and scale QD chains of 1d 
split-gate transistors by combining them into 2d arrays with sparse 
connectivity.
For this purpose, we introduce a new hardware junction which enables coupling 
between horizontally and vertically oriented arrays of split-gate transistors 
via spin shuttling~\cite{Yoneda2021}.
Given these hardware elements, we propose a modular 2d spin-qubit architecture 
with unit cells consisting of diagonally-oriented squares where 
nanowires form the edges and junctions the corners of a square.
We then investigate compilation strategies for the proposed 2d architecture and 
demonstrate a square-root scaling of the compilation overhead, which outperforms 
the linear overhead of 1d devices, not only asymptotically, but also down to 
its smallest building block which is likely to be investigated first in future 
experiments.
This allows for balancing the trade-off between compilation overhead and 
creating space in between the qubit modules which can be beneficial for 
colocation of classical control and readout electronics~\cite{Vandersypen2017, 
Boter2019, Ruffino2021a} and alleviating contact routing 
issues~\cite{Veldhorst2017, Li2018, Schaal2019, Pauka2021}.
Overall, this proposal should provide a compelling path to scaling, which could be manufactured with industrial CMOS processes.

\section{Results}

\subsection{Topology of the proposed architecture}
\begin{figure}
   \includegraphics[width=0.5\textwidth]{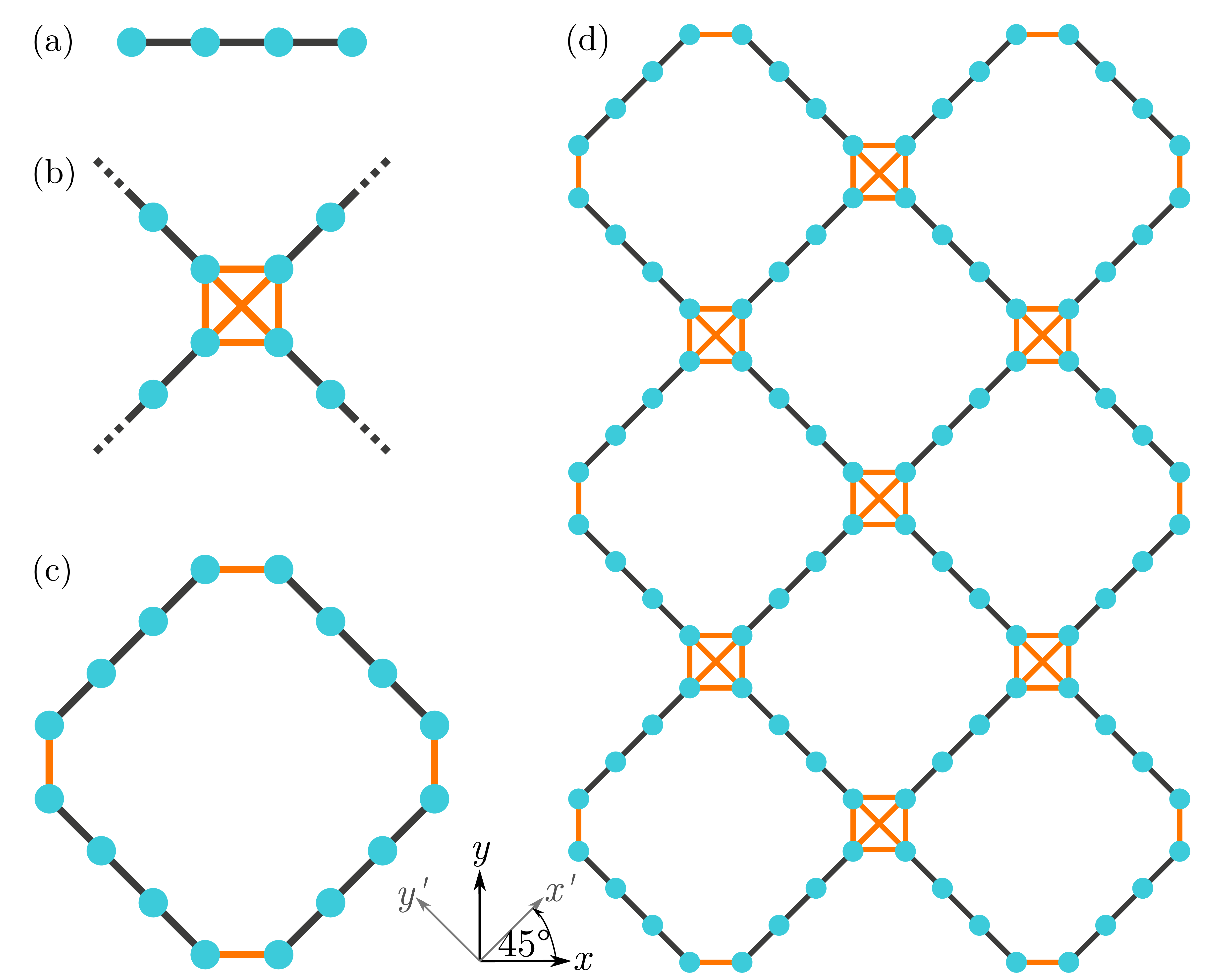}
   \caption{Hardware topology of the proposed architecture. (a) Linear segment 
of $m=4$ qubits (dots) connected by two-qubit operations (lines), representing 
split-gate transistors along a nanowire. (b) Hardware junction joining four 
linear qubit segments. Orange lines indicate two-qubit operations combined with 
spin shuttling. (c) Unit cell of the proposed architecture with $m=4$ qubits in 
each segment. (d) Hardware graph of the proposed architecture with $\dx=2$ and 
$\dy=3$ unit cells in x and y direction respectively.}
   \label{fig:sparsedevice}
\end{figure}

{\label{sec:topology}}
We start by introducing the hardware topology of the proposed architecture, 
before describing the detailed embodiment in silicon and discussing compilation 
strategies in subsequent sections.
The core element of the architecture is a bilinear array of QDs using
CMOS split-gate nanowire transistors.
This gives rise to a 1d arrangement of $m$ qubits along a line, in which neighboring 
qubits can be involved in two-qubit operations as visualized in 
Fig.~\ref{fig:sparsedevice}(a), where the $m=4$ dots indicate qubits and the black 
lines indicate possible two-qubit operations among neighbors.
An additional building block is provided by a junction element which can join 
up 
to four linear segments in a perpendicular manner, as depicted in Fig. 
\ref{fig:sparsedevice}(b).
In principle, a two-qubit operation can be executed between any two qubits 
involved in the junction as indicated by an orange line connecting the qubits.
As explained in more detail in the following section, each two-qubit operation 
across the junction requires 6 additional spin shuttling steps.
We consider the regime in which these coherent shuttling steps are fast, invoking negligible 
overhead as compared to the two-qubit operation.
We note that each qubit can only be involved in a single two-qubit operation 
at 
a time.

The unit cell of the proposed architecture is then provided by squares, with 
linear segments along the edges and junctions at the corners of each square, 
as in 
Fig.~\ref{fig:sparsedevice}(c).
To form the complete architecture, we join $\dx$ squares in one direction and
$\dy$ in the perpendicular direction.  For convenience, we
define the two directions as $x$ and $y$. The complete architecture and $x$
and $y$ directions are show in Fig.~\ref{fig:sparsedevice}(d).
%
Since each square contains $4m$ qubits and the device consists of $\dx\dy$ 
squares, each device contains
\begin{equation}
 \label{eq:NumQubits}
    \nq = 4m\dx\dy
\end{equation}
qubits.
We note that a crucial feature of the proposed architecture is the tilted 
orientation of squares at an angle of 45$^{\circ}$ in each unit cell, which  
minimizes the compilation overhead as will be explained later on.

\subsection{Architecture embodiment in silicon}{\label{sec:embodiment}}

In the following, we introduce the embodiment of the proposed architecture in 
silicon and the corresponding control schemes in more detail.
We focus on an implementation based on electron spins hosted in gate-defined QDs, but note that a similar proposal can be described for hole 
spin qubits.
Readers interested only in the compilation techniques are advised to skip ahead.

\begin{figure*}[t]
	\centering
		\includegraphics{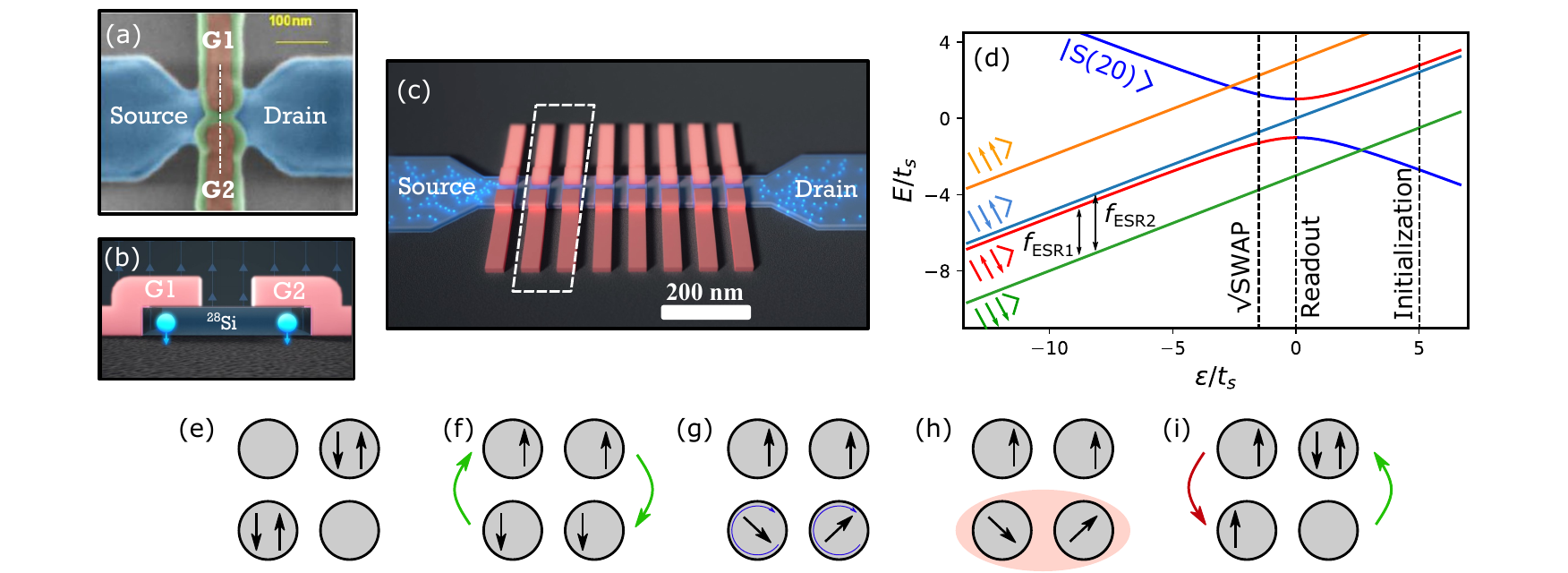}
	\caption{Split gate submodules. (a) Top view of a single split-gate 
nanowire transistor, with highly-doped source and drain ohmic contacts in blue, 
metallic split gates in orange, Si$_3$N$_4$ spacer in green and buried oxide in 
black. (b) Schematic cross section of the split-gate transistor along the dashed line in 
panel (a). The gates and silicon nanowire are isolated by the gate oxide. The 
blue 
arrowed spheres represent schematically the location of the spins. The vertical 
arrows represent the magnetic field lines. (c) An 8 split-gate 1d submodule. We highlight a $2 \times 2$ subset.  
(d) 
Energy spectrum versus energy detuning of the two-spin system in a
tunnel-coupled DQD. Both energy and detuning are normalized by the 
$\left|\uparrow\downarrow\right\rangle$-$\left|S(20)\right\rangle$ tunnel
coupling, $t_\text{s}$. The diagram is simulated using a 5\% g-factor difference 
and an average Zeeman energy of three times $t_\text{s}$. (e-i) Operation of a $2 \times 2$ QD subset. Initialization (e-f), one-qubit gates (g), two-qubit exchange interaction (h) and readout via Pauli spin blockade (i).
}
	\label{Fig:Nanowire}
\end{figure*}

\paragraph{Split-gate submodules}
The principal building block of our proposal is the split-gate 
transistor~\cite{Roche2012, Lundberg2020, Ansaloni2020, CirianoTejel2021}.
It consists of an undoped silicon-on-insulator nanowire of height $h$ and width 
$w$ (typically 10~nm and 60~nm, respectively).
The central part of the nanowire is gated by two metallic surface electrodes of length $l_\text{g}$ (typically 40-60 nm) 
which are isolated from the channel by the gate oxide (SiO$_2$) of thickness 
$t$ 
(typically ~6 nm); see Fig.~\ref{Fig:Nanowire}(a).
The gate stack is typically formed by 5~nm of TiN followed by 50~nm of 
polycrystalline silicon.
The rectangular cross-section of the channel, as seen in 
Fig.~\ref{Fig:Nanowire}(b), is covered by the pair of split gates which are 
separated by a face-to-face distance $S_\text{gg}$ ($\approx$ 30~nm) and enable 
local electrostatic control of the nanowire.
At deep cryogenic temperatures, when positive voltages are applied to the 
gates, 
few-electron QDs form in the top-most corners of the device due to the 
corner 
effect~\cite{Voisin2014, Ibberson2018, Ibberson2021}, as shown in Fig.~\ref{Fig:Nanowire}(b).
Charges can be drawn into the QDs from charge reservoirs formed of highly-doped 
silicon located at each side of the split gate. 

The qubit that we consider here is the spin of a single electron (or hole) 
confined to one of the two corner dots in each split-gate transistor.
The other spin is used as an ancilla for readout as described later on.
To define the spin quantization axis, the structure is placed in a magnetic 
field.
The architecture can be scaled-up by fabricating a series of split-gate 
transistors placed along the axis of the silicon nanowire, as depicted in 
Fig.~\ref{Fig:Nanowire}(c).
The split-gate edge-to-edge separation $S_\text{vv}$ ($\approx$ 40-60 nm) is set to enable 
sizable 
exchange coupling between spins, which we use to generate two-qubit interactions.
Overall, the module results in a bilinear array of QDs.
Given that the two QDs of each split-gate transistor encode one qubit -- one dot contains a qubit spin and the other an ancilla spin for readout -- the structure embodies a one-dimensional chain of silicon spin qubits of length $m$, as described in the previous section.

Next, we explain control, readout and initialization of the 1d modules in detail. 
We base our explanation on the energy spectrum of the coupled two-spin system in 
the single spin basis as a function of the QD energy detuning, $\epsilon$, and 
a finite magnetic field; see Fig~\ref{Fig:Nanowire}(d). At large positive detuning with respect to the (11)-(20) charge hybridization point, 
the ground state of the system corresponds to the intradot singlet, 
$\left|S(20)\right\rangle$. Here the ($nm$) notation refers to the charge 
distribution among the two QDs, i.e. dots are occupied with $n$ and $m$ charges, respectively. At 
negative detuning, the ground state of the system is the (11) charge 
configuration whose spin degeneracy is broken by the external magnetic field. We consider the QDs have a tunnel coupling energy $t_\text{s}$ and present different $g$-factors due to the variability of the Si/SiO$_2$ interface~\cite{Huang2017}, which further breaks the degeneracy of the 
$\left|\uparrow\downarrow\right\rangle$ and 
$\left|\downarrow\uparrow\right\rangle$ states. Here, the first spin refers to the spin with the lower $g$-factor (see the Hamiltonian in Appendix A). Although not strictly necessary for the operation of the processor, we consider that larger Zeeman energy differences exist across the nanowire than along the nanowire, which could be achieved by creating an asymmetry in the gate structure. Using magnetic materials on the gate stack on one side of the split or placing the splits offset with respect to the nanowire axis~\cite{Ibberson2021} may produce a difference in the Meissner effect at the location of the QDs. We use the energy spectrum to describe processes involving QDs within a split-gate and QDs on the same corner of the nanowire in different splits.   

\paragraph{Initialization}

We start with the 1d module loaded with one charge in each QD. Charges can be 
drawn in from reservoirs at the periphery following established 
methods~\cite{Volk2019}. To initialize to a known spin state, QDs in each split 
gate are positively detuned until the system relaxes to the 
$\left|S(20)\right\rangle$ state (Fig.~\ref{Fig:Nanowire}(e)). Then, the system 
is pulsed towards negative detuning adiabatically with respect to the 
$\left|\uparrow\downarrow\right\rangle$-$\left|S(20)\right\rangle$ coupling, to 
initialize the QDs in the splits in the $\left|\uparrow\downarrow\right\rangle$ 
state (Fig.~\ref{Fig:Nanowire}(f)).

\begin{figure*}
	\centering
		\includegraphics{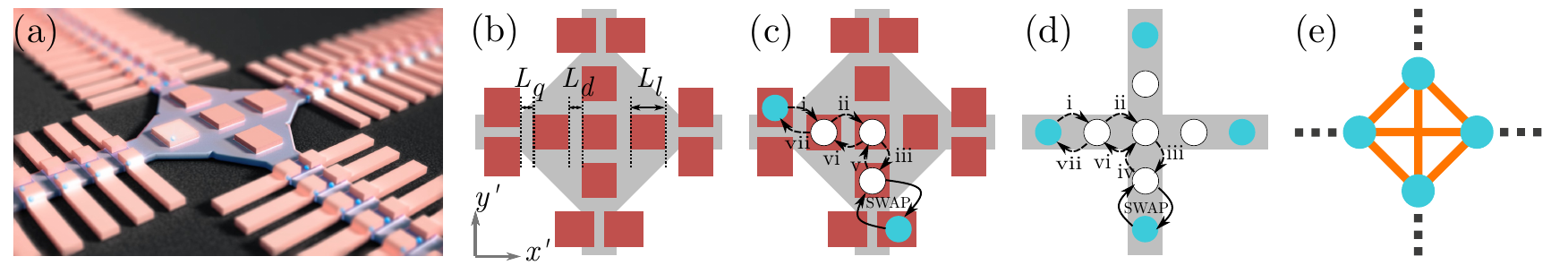}
	\caption{Junction and shuttling sequence. (a) Schematic diagonal view of 
the 
junction connecting two $x'$ and two $y'$ submodules. (b) Top view of the 
junction. (c) Shuttling sequence from $x'$ submodule at the left to $y'$ 
submodule at the bottom. Arrows indicate three consecutive spin shuttling 
operations (i-iii), followed by a \textsc{Swap} operation, and another three consecutive 
spin shuttling operations (v-vii). (d) Same as (c), abstracted. (e) Hardware 
topology of the junction. Qubits (dots) of $x'$ and $y'$ submodules are 
connected via six shuttling and one \textsc{Swap} operation (lines).
}
	\label{fig2}
\end{figure*}

\paragraph{Single-qubit operations}
Electron spins in isotopically purified silicon present long 
coherence times ($T_{2}^{*} > 100$~$\mu$s and $T_2=28$~ms)~\cite{Veldhorst2014}.
This enables high control fidelity via electron spin resonance (ESR) techniques~\cite{Pla2013, Veldhorst2015}.
Manipulation occurs for QDs on the same side of the nanowire which are now initialized to the $\left|\downarrow\downarrow\right\rangle$ state. Individual spin transitions can be addressed with an oscillatory magnetic field at frequency $f_{\text{ESR},i}$ if in resonance with the Zeeman energy of the relevant spin transition (Fig.~\ref{Fig:Nanowire}(g)).
Typical Rabi frequencies achieved with this method are of the order of 
1~MHz with fidelities in excess of $F > 99$\%~\cite{Yang2019}.
The ESR implementation for the proposed architecture requires placing the structure inside a broadband 3D microwave cavity~\cite{Vahapoglu2021, vahapoglu2021b}.

%
%
The excitation is applied globally, and the qubits would 
be tuned in and out of resonance making use of the Stark shift using local voltage 
pulses at the qubit gate~\cite{Laucht2015}.
ESR allows two axis control (X and Y gates) by controlling the duration of the gate voltage pulses and the phase of the microwave excitation. 
Implementing the proposed architecture with hole-spin qubits, on the other hand, 
would result in shorter coherence times ($T_{2}^{*} > 
250$~ns)~\cite{Maurand2016, Camenzind2021}.
However, the fact that holes exhibit larger spin-orbit coupling when compared to 
electron spins would allow for all-electrical control via the QD gates at relatively fast Rabi 
frequencies of up to 150~MHz via electric-dipole spin resonance 
(EDSR)~\cite{Bosco2021}.
We note that full control over the Bloch sphere of hole spins via X and Y 
rotations has been recently demonstrated with fidelities in excess of 99\%~\cite{Camenzind2021}.

\paragraph{Two-qubit operations} 
We propose engineering two-qubit gates between spins on the same corner of the 
nanowire by means of the spin-spin exchange interaction (Fig.~\ref{Fig:Nanowire}(h)).
The exchange interaction can be modulated electrostatically by applying a 
differential mode voltage on the two relevant neighboring gates that brings 
the system close to $\epsilon=0$; see Fig~\ref{Fig:Nanowire}(d) and Fig~\ref{Fig:Nanowire}(h).
In the limit where the differential mode voltage pulse increases the exchange 
coupling beyond the Zeeman energy difference between spins, a 
$\sqrt{\textsc{Swap}}$ or $\textsc{Swap}$ gate can be implemented by timing the duration and depth of the interaction pulse~\cite{Maune2012, He2019}. 
Alternatively, when the size of the modulation is smaller, such that the 
exchange coupling strength remains below the Zeeman energy difference, a CPhase 
gate could be implemented~\cite{Veldhorst2015} with fidelity above fault-tolerant thresholds~\cite{xue2021a,noiri2021,mills2021}. In this article, we focus on the former gate but note that the \textsc{Swap} gate can be synthesized from a combination of CPhase and single qubit gates.  

\paragraph{Readout}
Spin readout is based on spin-dependent tunneling from the (11) to the (20) charge configurations, i.e. Pauli spin blockade~\cite{Ono2002}. More particularly, the state $\left|\uparrow\downarrow\right\rangle$ is allowed to tunnel to the $\left|S(20)\right\rangle$ state, whereas all remaining two particle spins states are blocked~\cite{Veldhorst2017,Zhao2019}; see Fig~\ref{Fig:Nanowire}(d) and (i).
To detect this tunneling process, we suggest using dispersive readout~\cite{GonzalezZalba2015}.
One of the split gates is connected to a lumped-element electrical resonator 
which is driven at its natural frequency, $f_0$.
%
%
At $\epsilon=0$, cyclic tunneling between the 
$\left|\uparrow\downarrow\right\rangle$ and $\left|S(20)\right\rangle$, driven 
by the oscillatory resonator voltage, manifests in an additional quantum capacitance that loads the 
resonator producing a spin-dependent frequency shift that can be readily detected with standard methods~\cite{Mizuta2017,Pakkiam2018, West2019, Zheng2019, Crippa2019}.
\paragraph{Junctions}
Next, we propose a new hardware junction which allows for coupling 1d submodules 
of split-gate transistors.
The junction consists of etched silicon-on-insulator in quadrangular form, as 
can be seen in the central region of Fig.~\ref{fig2}(a).
The exact shape of the junction can vary to accommodate different levels of 
interconnection.
Here, we present a square junction that enables connecting up to four 
submodules at an angle of 90$^{\circ}$, 180$^{\circ}$, or 270$^{\circ}$.
On top of the junction, we place a series of metallic gates with the same gate 
stack as the split-gate transistors.
We propose a five-gate structure arranged in cross geometry -- one central square
gate flanked by four square gates of the same footprint.
The characteristic dimensions of the junction are indicated in 
Fig.~\ref{fig2}(b), with submodule-to-edge-gate separation in $x'$($y'$) 
directions $L_\text{q}$ ($\approx 40$~nm), the gate length in the $x'$($y'$) 
directions $L_\text{l}$ ($\approx 80$~nm), and a junction gate-to-gate 
separation $L_\text{d}$ ($\approx 40$~nm).
The junction is invariant under rotations by 90$^{\circ}$.

The purpose of the junction is to create a shuttling path for 
electrons~\cite{Mills2019, Yoneda2021} at the edges of the 1d submodules to be 
moved around the junction in the $x'$ and $y'$ directions on demand using the 
appropriate gate voltages sequence.
These electrons can be moved to be in exchange coupling proximity with the 
electrons at the edge of another submodule where a two-qubit gate will be 
performed.
We illustrate the operation of the junction using an $x'y'$ coupling example in 
Fig.~\ref{fig2}(c): 
(i) Shuttle the electron under the top-rightmost gate in the left module 
$x'$, to the left gate in the junction by applying a differential voltage 
between the two gates~\cite{Yoneda2021}.
(ii) Shuttle from the left gate to the central gate of the junction. 
(iii) Shuttle from the central gate to the bottom gate of the junction.
(iv) Implement a two-qubit gate between the electron under the bottom 
gate in the junction and the electron under the left-topmost gate in the $y'$ 
module, as described above.
(v-vii) Shuttle the electron back following the reverse process. 
Finally, the abstracted hardware topology and the same shuttling sequence are presented in Fig.~\ref{fig2}(d). Although in this work we consider two-qubit interactions between spins confined to different types of QDs (a corner QD in a submodule and a planar QD in the junction), the shuttling process above can be modified to produce two-qubit interactions between planar QDs which have been demonstrated~\cite{Maune2012, Veldhorst2015,xue2021a,noiri2021} .  
We consider the regime in which spin shuttling operations are performed coherently and much faster 
than the two-qubit operation in step (iv). In this case, the overhead is negligible and 
we use the simplified hardware topology given in Fig.~\ref{fig2}(e). See Appendix B for a discussion of the different overhead regimes.  

\subsection{Compilation methods}

We now present compilation methods for deploying quantum algorithms on the 
proposed architecture, following standard approaches.
We assume that algorithms are given by a quantum circuit which
consists of initialization, followed by a sequence of one- and two-qubit 
operations, and readout, suitable for execution on a fully-connected device.
While most of these operations are readily available on the proposed architecture, the 
sparse connectivity will prohibit execution of two-qubit gates between 
arbitrary pairs of qubits that are not directly connected by an edge of the hardware 
graph.
To address this issue, compilation methods re-express a given quantum circuit 
through an equivalent circuit readily amenable to the sparse hardware topology.
More specifically, before executing a given two-qubit gate, one generally 
applies a sequence of \textsc{Swap} gates to shuttle the relevant qubits along 
the hardware graph until they reach neighboring positions.
Once neighboring positions are reached, the two-qubit gate is executed and the 
sequence of \textsc{Swap} gates is reverted to promote the relevant qubits 
back to their original position.
In applying this compilation strategy, the resulting quantum circuit accumulates 
a compilation overhead characterized by the number of additional \textsc{Swap} 
gates, $\ns$, and the increased circuit depth, $\nd$.
Both quantities represent important quality metrics for the compilation which 
should be minimized to avoid additional decoherence and infidelities introduced 
through the additional \textsc{Swap} gates.
We note that the proposed architecture uses two $\sqrt{\textsc{Swap}}$ gates to 
implement the \textsc{Swap} operation.

In what follows, we focus on two major compilation scenarios:\
(I) The case of moving two arbitrary qubits together along the hardware graph.
This covers the general case where each two-qubit operation is addressed 
individually and usually gives reasonable estimates for the compilation 
overhead.
(II) The case of rearranging qubits in an arbitrary permutation.
This compilation method is useful for variational algorithms \cite{Cerezo:2021aa} from chemistry \cite{Peruzzo:2014aa, fedorov2021vqe} 
and finance \cite{ORUS} which repeatedly execute some number of up to 
$\left\lfloor\nq/2\right\rfloor$ two-qubit operations in parallel.
For such algorithms, the compilation must efficiently permute qubits into 
configurations which allow for executing the relevant two-qubit operations in 
parallel, even on the sparse hardware topology.
To implement such permutations efficiently, one repeatedly applies a layer of 
\textsc{Swap} operations during which up to $\left\lfloor\nq/2\right\rfloor$ 
\textsc{Swap} gates are executed in parallel.
Ultimately, this results in significantly lower circuit depth and reduced gate 
count than simply addressing each two-qubit operation individually.

\begin{figure}
  \includegraphics[width=0.5\textwidth]{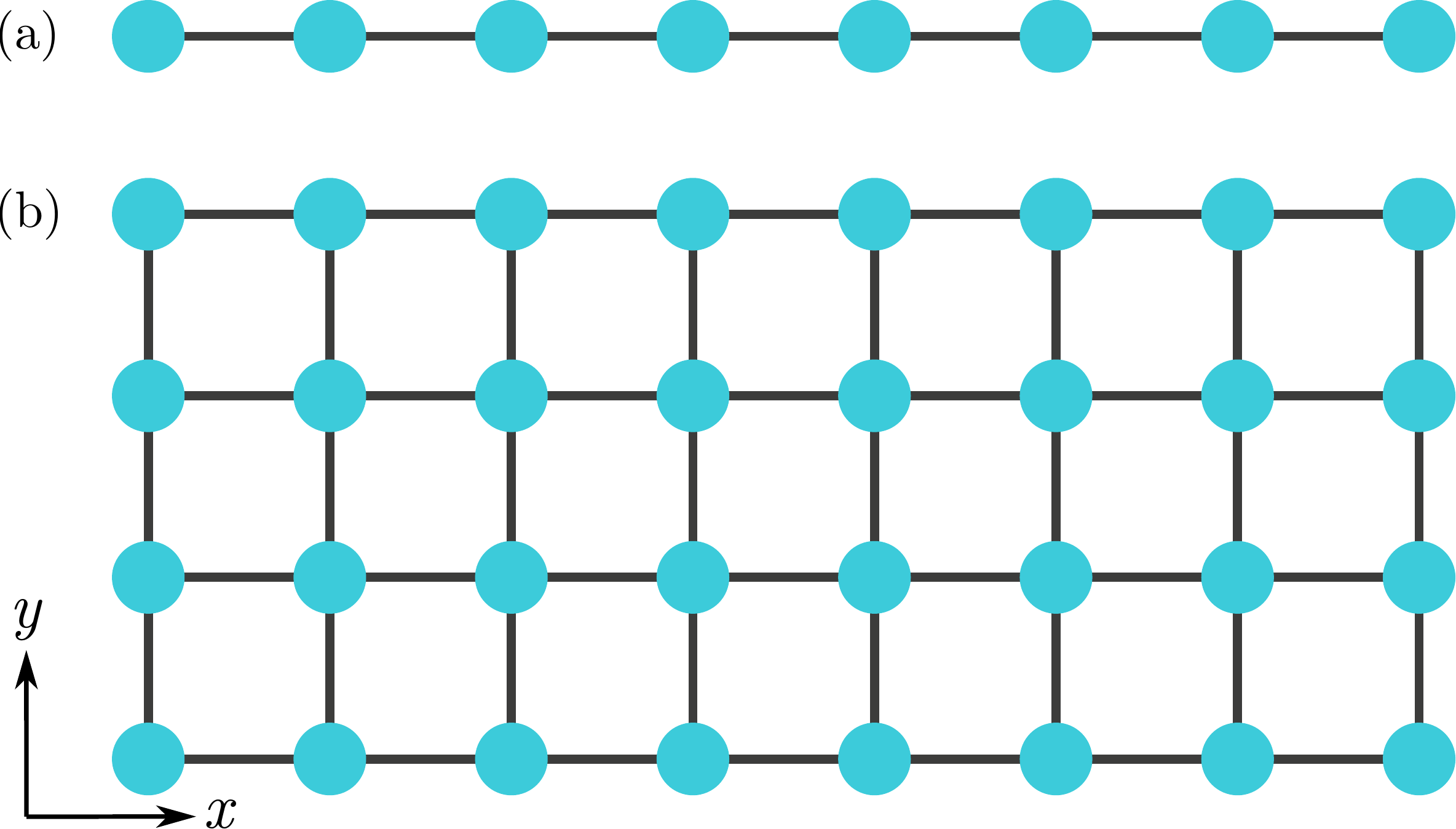}
  \caption{(a) Linear layout with $\nq=8$ and (b) rectangular layout with 
nearest-neighbor connectivity with $(\lx,\ly)=(8,4)$. Qubits are depicted as 
circles and lines indicate pairs between which a two-qubit operation is 
feasible.}
  \label{fig:otherdevices}
\end{figure}

Finally, we compare our compilation methods to two important limiting cases -- 
a 1d device which could be fabricated by joining nanowire submodules along one 
dimension, as depicted in Fig.~\ref{fig:otherdevices}(a), and 
a 2d rectangular device consisting of $l_{x}$ by $l_{y}$ qubits with 
nearest-neighbor connectivity, as depicted in Fig.~\ref{fig:otherdevices}(b).

\subsubsection{Case I -- Moving two qubits together along the shortest path}

\begin{figure*}
    \centering
    \includegraphics[width=\textwidth]{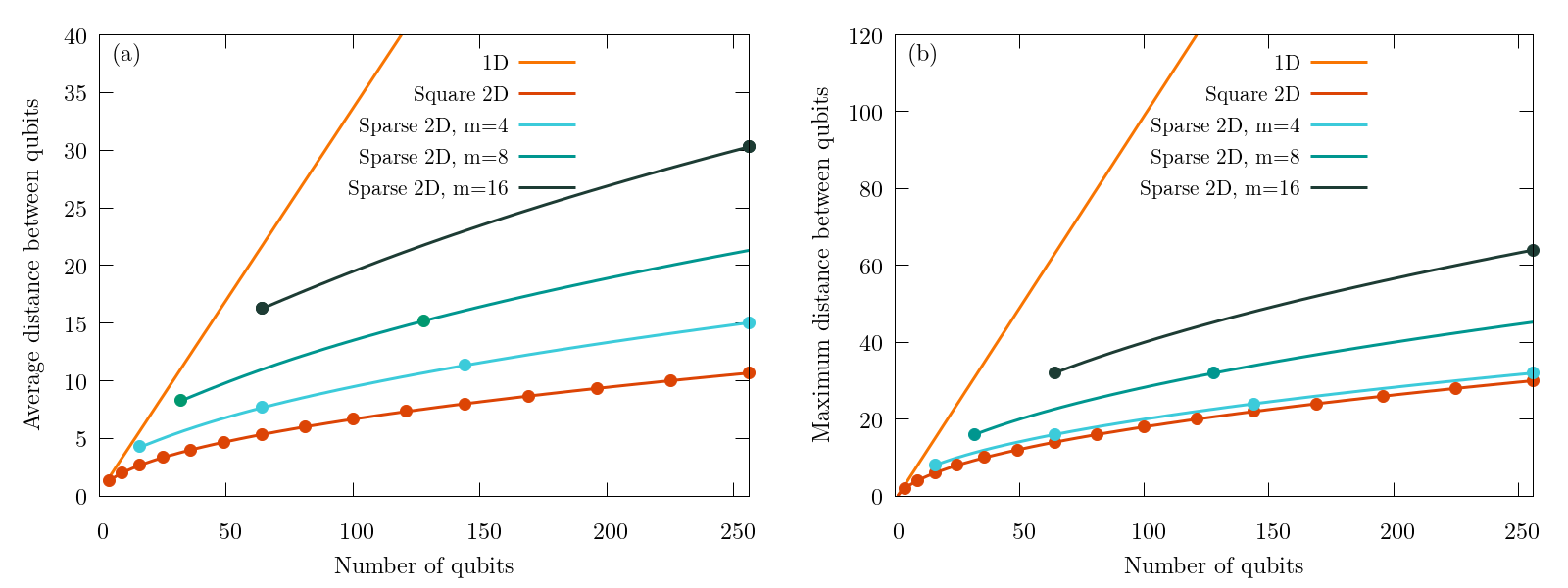}
    \caption{Compilation overhead expressed via (a) average and (b) maximum 
      shortest path as a function of the number of qubits for different layouts
      -- a (1D) linear layout, a (2D) rectangular device with nearest 
neighbor connectivity with $l_x=l_y$ and the proposed architecture in a square 
arrangement $\dx=\dy$ with $m=4$, 8, and 16. Lines indicate the 
scaling and dots indicate qubit numbers which can be realized by an actual 
device.
}
    \label{fig:avemaxdist}
\end{figure*}
\paragraph{Algorithm}
The compilation method for case I requires iterating over all two-qubit operations
of a given input circuit and, for each pair of qubits involved in a two-qubit operation,
(a) determining the shortest path connecting the two qubits and subsequently
(b) executing \textsc{Swap} gates to connect the qubits along the shortest path, executing 
the two-qubit operation and shuttling qubits back to their original position.
Finding the shortest path, i.e, the distance, between a pair of qubits can
efficiently be implemented in polynomial time using, e.g., Dijkstra's algorithm 
\cite{Cor2009}.

\paragraph{Overhead}
The compilation overhead of this algorithm is determined by the length of the 
shortest path, $l$, connecting the pair of qubits involved in each two-qubit 
operation along the hardware graph.
In particular, the overhead of \textsc{Swap} gates accumulated per two-qubit 
gate is given by $\ns = l-1$ while the increased circuit depth per two-qubit operation will be given by 
$\nd = \left \lceil \frac{l-1}{2} \right \rceil$,
assuming that \textsc{Swap} gates on different pairs of qubits can be applied 
simultaneously.

\paragraph{Device specific overhead}
To compare further the compilation overhead for different devices, we compute the 
average ($\bar{l}$) and maximal ($l_{\mathrm{max}}$) distances between qubit 
pairs in a given topology.
For the linear layout with $\nq$ qubits, we have
\begin{equation}
    \bar{l}_{\linear} = \frac{1}{3}(\nq+1), \quad l_{\linear,\mathrm{max}} = 
\nq-1.
\end{equation}
For a rectangular grid of $\lx$ by $\ly$ qubits, we find that $\bar{l}_{\rect} 
= \frac{1}{3}(\lx+\ly)$ and $l_{\rect, \mathrm{max}} = \lx+\ly - 2$ and 
specifically for the square grid with $\lx = \ly = \sqrt{\nq}$ we have
\begin{equation}
\bar{l}_{\rect}=\frac{2}{3}\sqrt{\nq}, \quad 
l_{\rect,\mathrm{max}}=2(\sqrt{\nq} - 1).
\end{equation}
Finally, considering the proposed architecture for $\dx=\dy$, the average 
and maximal qubit distances are
\begin{align} \label{eq:lbarmd}
\bar{l}_{\twod} = & \frac{1}{45(\nq-1)}\sqrt{\frac{m}{\nq}}\left[21 \nq^{2} + 5 
                \left (m + \frac{2}{m} \right )\nq \right. \\ 
                & \left. - 15 \left (\sqrt{m^{3}} + \frac{2}{\sqrt{m}}\right 
                )\sqrt{\nq} + 34 m^{2} + 20 \right], \nonumber \\
                \simeq & \frac{7}{15} \sqrt{m\nq}, \nonumber\\ 
    \bar{l}_{\twod,\max} =& \sqrt{m\nq}. \nonumber
\end{align}
For details of the derivation, see Appendix C.

\begin{figure*}
  \includegraphics[width=\textwidth]{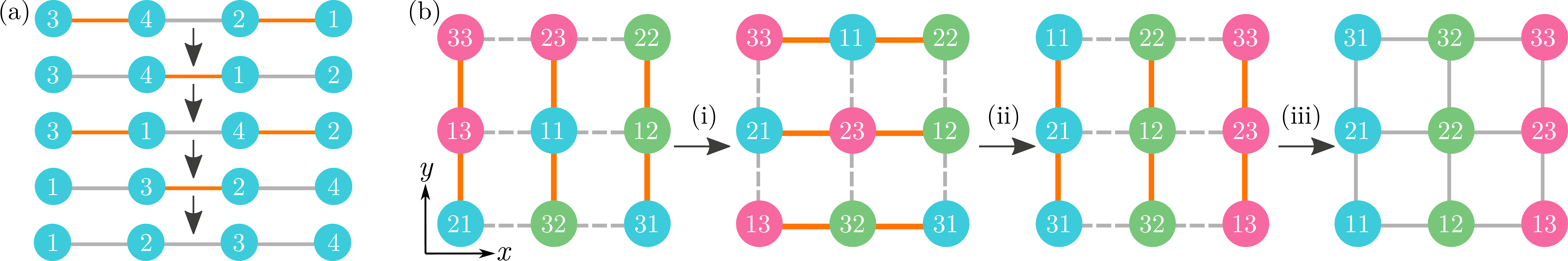}
  \caption{Qubit permutations on (a) linear and (b) rectangular devices.
(a) Parallel neighbor sort for a linear device with $\nq=4$ qubits using 4 
consecutive layers of \textsc{Swap}s. Qubits (circles) are labeled by final 
position. Orange lines indicate qubit pairs which are compared at a given 
iteration and swapped, if in the wrong order.
(b) Permuting $\nq=9$ qubits (circles) in a square layout. Labels indicate the final 
row and column. The permutation is implemented by consecutively using parallel 
neighbor sort along (i) columns, (ii) rows and (iii) columns, as highlighted by 
orange lines.
}
  \label{fig:2dswap}
\end{figure*}
We visualize the aforementioned compilation overheads expressed via mean and 
maximal qubit distances in Figs.~\ref{fig:avemaxdist}(a) and 
\ref{fig:avemaxdist}(b) as a function of increasing qubit number.
These figures illustrate several important properties, which we comment on 
in the following.
The linear configuration (labeled 1D) clearly exhibits a linear scaling of the 
compilation overhead, while the rectangular device with nearest neighbor 
connectivity (labeled 2D) exhibits a square-root scaling.
The compilation overhead of the proposed architecture inherits the favourable 
square-root scaling of 2d hardware topologies, making the proposed architecture 
favorable over 1d devices.
Interestingly, the compilation overhead of the proposed device never exceeds the 
compilation overhead of the linear architecture, even for the smallest devices 
$\dx=\dy=1,2,3,...$.
This is useful as first experimental realizations of the proposed architecture 
would start from small prototypes.
Finally, for increasing sparsity $m=4, 8, 16, ...$, the compilation overhead of 
the proposed architecture does increase; however, it never exceeds the overhead 
of the linear device.
This allows for balancing the trade-off between compilation overhead and 
creating space in between the qubit modules which can be beneficial for 
colocation of classical support electronics.

\subsubsection{Case II -- Rearranging qubits in arbitrary permutations}
\label{sec:methodlayer}

We now consider the cost of permuting all qubits at once, making use of sorting 
networks~\cite{hirata2011efficient,Beals_2013}.
We begin by recalling qubit permutations on linear and rectangular devices 
\cite{Steiger_2019, brierley2015efficient} as these underlie the compilation 
method for the proposed sparse architecture.

\paragraph{Parallel neighbor sort for 1d}
We first consider permutations on 1d linear devices with $\nq$ qubits  
using parallel neighbor sort~\cite{habermann1972parallel}, as depicted in Fig.~\ref{fig:2dswap}(a).
First, each node of the hardware graph is assigned an index $v=1,...,\nq$ in 
ascending order, and each qubit is labeled by its final position.
Consecutive layers of \textsc{Swap} operations are then applied in 
up to $\nq$ steps.
For odd steps $1,3,...$, qubits on node pairs $(v,u)\in\{(1,2), (3,4), ...\}$
are compared and a \textsc{Swap} operation is applied if the final destination
of qubit $v$ is larger than the final destination of qubit $u$.
For even steps $2,4,...$, the same process occurs for qubits in  
node pairs $(v,u)\in\{(2,3), (4,5), ...\}$.
With this method, any permutation can be implemented in a maximum of $\nq$ 
layers leading to an increase in the circuit depth of
\begin{equation}
 \nd = \nq.
\end{equation}
The maximum number of \textsc{Swap} operations needed is
\begin{equation}
 \ns = \frac{\nq (\nq-1)}{2}.
\end{equation}
An alternative is to use bubble sort or insertion sort, but the maximum depth
increases to $2\nq-3$~\cite{hirata2011efficient,Beals_2013}.

\paragraph{Qubit permutation for the rectangular device}
Next, we consider permutations on a rectangular device of $\nq= \lx\ly$ qubits with nearest neighbor 
connectivity.
We follow the algorithm of Ref.~\cite{Steiger_2019}, which consists of the 
following three steps:
(i) Rearrange the qubits in each column using parallel neighbor sort such that 
each row contains exactly one qubit with final destination in each column $1, 2, 3, 
..., \lx$.
(ii) Rearrange the qubits in each row using parallel neighbor sort such that 
all qubits are in the correct column.
(iii) Rearrange the qubits in each column using parallel neighbor sort such 
that each qubit is in the correct final location.
We note that column and row can also be interchanged in the above steps.
An example visualizing the method is shown in Fig.~\ref{fig:2dswap}(b).

Once step (i) provides an arrangement such that each row contains 
exactly one qubit with final destination in column $1, 2, 3, ..., \lx$, an 
implementation of steps (ii) and (iii) using parallel neighbor sort is simple.
The challenge is to see that an efficient implementation of step (i) is 
always possible.
This was shown in Refs.~\cite{brierley2015efficient,Steiger_2019} using Hall's 
matching theorem~\cite{hall1935representatives} and will be discussed in more detail in its adaption to the proposed 
architecture with sparse 2d connectivity below.
The overhead of the discussed method originates from consecutively 
using parallel neighbor sort on (i) columns (ii) rows and (iii) columns. This
results in a maximum of
\begin{equation}
  \nd = 2\ly + \lx
\end{equation}
layers of \textsc{Swap} gates and
\begin{equation}
  \ns = \frac{1}{2}\lx\ly(\lx+2\ly-3)
\end{equation}
total \textsc{Swap} gates.
Specifically, for the square $\ly=\lx=\sqrt{\nq}$, this gives
\begin{equation}
    \nd = 3\sqrt{\nq}
\end{equation}
and
\begin{equation}
  \ns = \frac{3}{2}\nq (\sqrt{\nq}-1).
\end{equation}

\begin{figure}[]
  \includegraphics[width=0.5\textwidth]{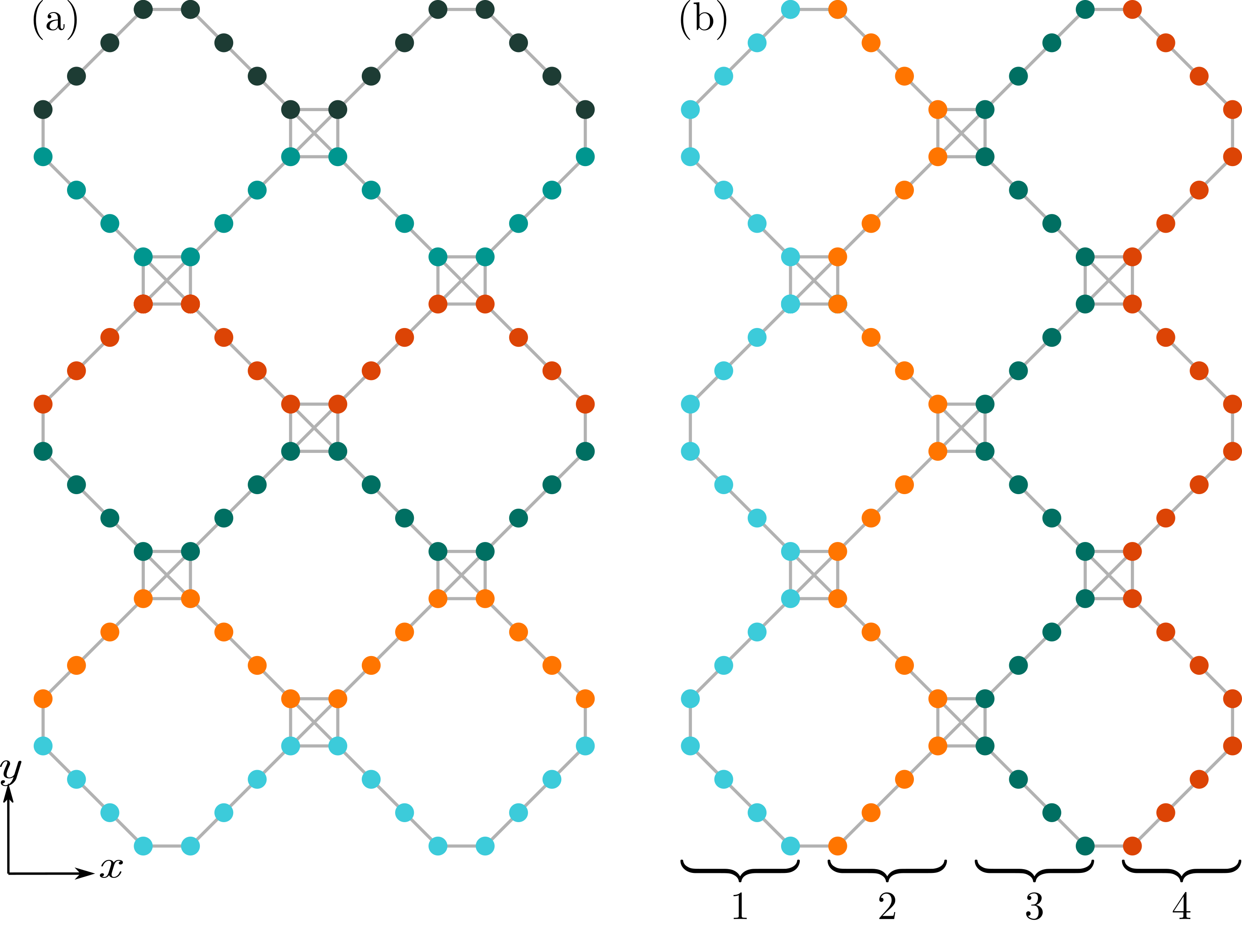}
  \caption{The generalized (a) rows and (b) columns of the proposed architecture 
with 
$m=4$, $d_{x}=2$ and $d_{y}=3$. Qubits of the same colour are in the same row 
or column. Lines indicate between which qubits a two-qubit gate can be applied.
}
  \label{fig:sparseconn2}
\end{figure}

\begin{figure}
  \centering
  \includegraphics[width=0.4\textwidth]{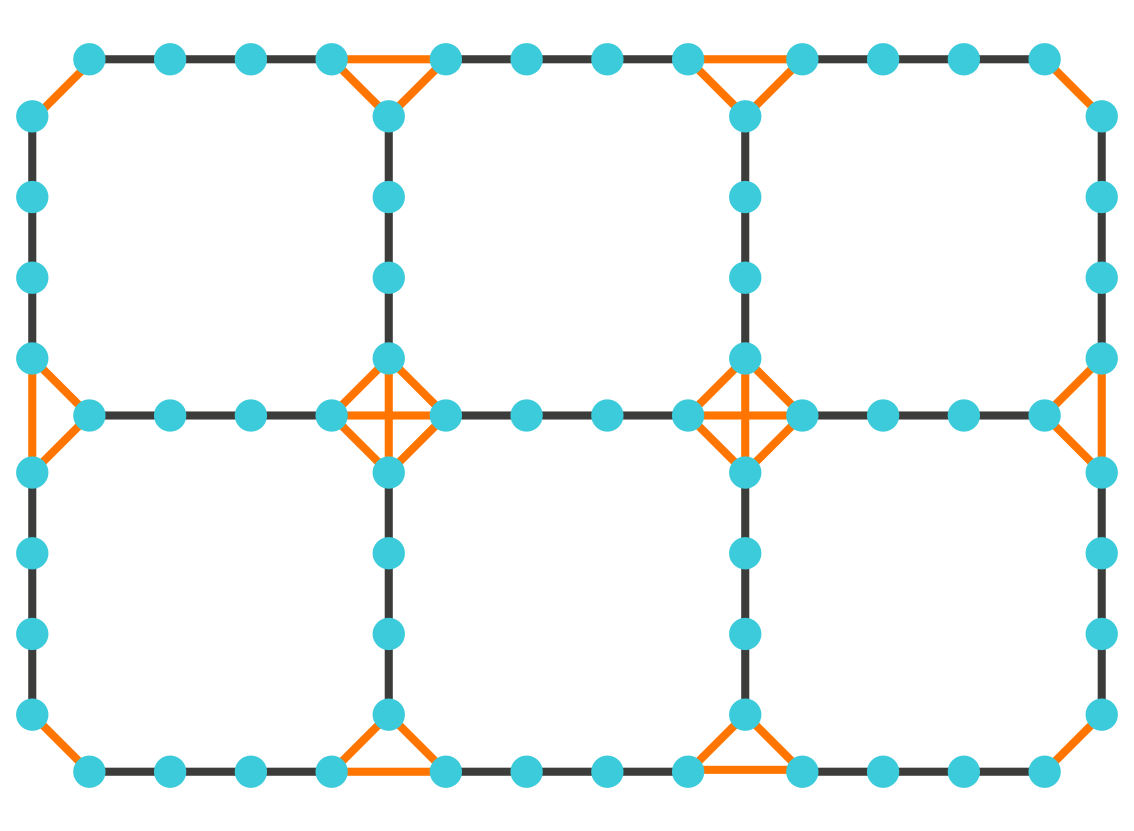}
  \caption{An alternative sparsely-connected device constructed from the 
linear segments and four-way junctions shown in 
figures~\ref{fig:sparsedevice}(a) 
and~\ref{fig:sparsedevice}(b) respectively. Qubits are represented by dots and 
lines 
indicate between which qubits a two-qubit gate can be performed. We note that 
this device does not 
lend itself to the definition of generalized rows and columns outlined in the 
text.}
  \label{fig:othersparsedevice}
\end{figure}

\paragraph{Generalized rows and columns}
We now extend the method of Ref.~\cite{Steiger_2019} to the proposed 
architecture with sparse 2d connectivity.
Our method is based on the definition of generalized rows and columns as 
depicted in Figs.~\ref{fig:sparseconn2}(a) and \ref{fig:sparseconn2}(b) 
respectively.
In essence, this results in $2d_{y}$ rows and $2d_{x}$ columns with $2md_{x}$ 
and $2md_{y}$ qubits respectively, where any given row and column share $m$ 
qubits.
Having defined these generalized rows and columns, we note that every qubit 
can conveniently be labeled by a combination of its column index, 
$a_x=1,...,2d_{x}$, and its vertical position, $y$. Equivalently, we could choose to label a qubit by its row index, $a_{y}$
and its horizontal position, $x$. The coordinates $x$ and $y$ can be seen in figure~\ref{fig:sparsedevice}(d).
We note that not all constructions of sparsely-connected devices from the linear segments
and junction shown in figures~\ref{fig:sparsedevice}(a) and \ref{fig:sparsedevice}(b)
respectively would have enabled such simple definitions of generalized rows and columns.
We show an alternative device construction in figure~\ref{fig:othersparsedevice}, for which equivalent
generalized rows and columns do not exist. It is also not possible to draw a single line
through all the qubits in such a device.

\begin{figure*}
  \includegraphics[width=\textwidth]{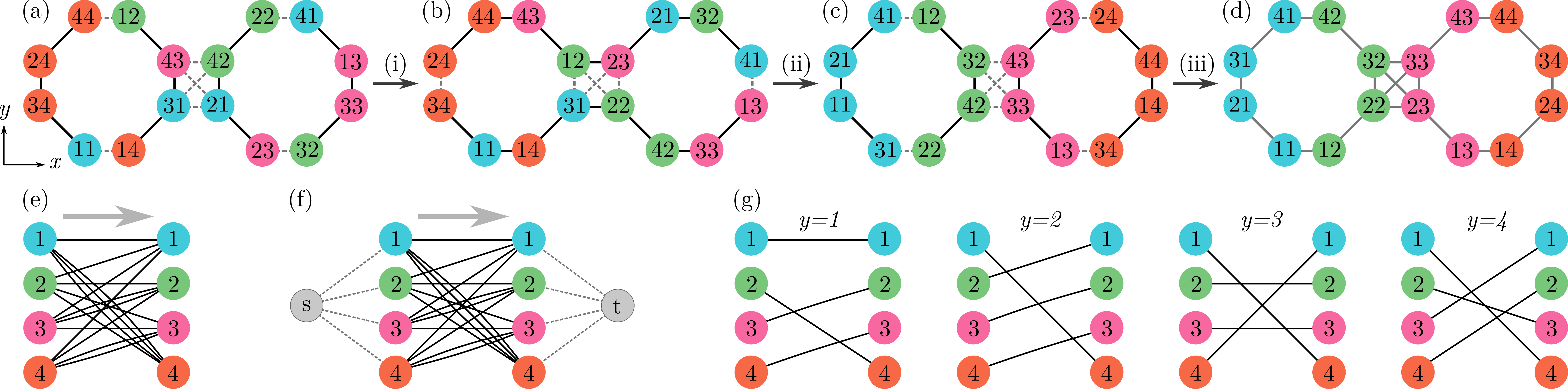}
  \caption{Qubit permutation on proposed architecture with $\dx=2$, $\dy=1$, 
and $m=2$.
(a-d) Qubits (circles) labeled by their final destination $(y, a_x)$ are 
permuted using parallel neighbor sort along generalized (i) columns, (ii) rows, 
and (iii) columns as indicated by black solid as opposed to gray dashed lines. 
(e) Bipartite graph corresponding to step (i). Nodes on the left and right 
indicate qubits with origin and final destination in columns 1, 2, 3, and 4 
respectively, while edges represent the corresponding qubit.
(f) Same as (e) with virtual nodes $s,t$.
(g) Matchings extracted from (e). The $y$th matching determines qubits routed 
to position $y$. An edge $(l,r)$ of $y$th matching indicates that a qubit of 
column $l$ destined for column $r$ is to be routed to position $y$.
}
  \label{fig:CompProposed}
\end{figure*}

\paragraph{Compilation for the proposed architecture}
With generalized rows and columns in place, we describe the corresponding 
compilation algorithm for implementing qubit permutations.
Each qubit, in its initial location, carries a label 
$(a_x,y)$, indicating its final position.
An arbitrary qubit permutation can then be implemented using the following three steps:
(i) Rearrange the qubits in each generalized column using parallel neighbor 
sort such that each set of qubits with fixed coordinate $y$ contains exactly one qubit 
with final destination in each column $a_x=1, 2, 3, ..., 2\dx$.
(ii) Rearrange the qubits in each generalized row using parallel neighbor sort 
such that all qubits are moved into the correct column according to their 
column index $a_x$.
(iii) Rearrange the qubits in each generalized column using parallel neighbor sort 
such that each qubit is in the correct $y$ location along the columns.
An example visualizing the method is given in Fig.~\ref{fig:CompProposed}.

\paragraph{Bipartite routing graph}
We note again that, once step (i) provides an arrangement such that each set of qubits 
with vertical position $y$ contains exactly one qubit with final destination in column 
$a_x=1, 2, 3, ..., 2\dx$, an implementation of steps (ii) and (iii) using 
parallel neighbor sort is straightforward.
The challenge is again to see that an efficient implementation of step (i) is 
always possible.
The procedure to achieve this is illustrated in Fig.~\ref{fig:CompProposed}(e-g) 
and shall now be explained in more detail.
To begin with, a bipartite graph of $4\dx$ nodes is constructed,
$l\in{1,...,2\dx}$ on the left and $2\dx$ nodes $r\in{1,...,2\dx}$ on the right.
Nodes on the left indicate columns in which a qubit is located initially.
Nodes on the right indicate columns to which a qubit should be routed.
To build the graph, one adds 
an edge $(l,r)$ to the bipartite graph for each qubit initially located in 
column $l$ and having final destination in column $r$.
An example is given in Fig.~\ref{fig:CompProposed}(e).
We note that, since we have $2m\dy$ qubits located in each column initially and 
since we will have $2m\dy$ qubits with final destination located in each 
column, the bipartite graph has $2m\dy$ edges incident to each node.
Since some qubits may originate and end up in the same column, the 
bipartite graph can have multiple edges connecting the same nodes.

\paragraph{Hall's matchings}
Next, to determine how qubits should be arranged along columns in step (i), one 
extracts $2m\dy$ perfect matchings $\mathcal{M}_{y}$ with $y=1,...,2m\dy$ from the 
bipartite graph.
A perfect matching $\mathcal{M}_{y}$ is a set of edges such that each node of the 
bipartite graph is connected to exactly one edge of the matching.
See Fig.~\ref{fig:CompProposed}(g) for examples showing one possible set of perfect
matchings for the bipartite graph in Fig.~\ref{fig:CompProposed}(e).
Finding $2m\dy$ perfect matchings for the given type of bipartite graph is 
always possible due to Hall's matching theorem~\cite{hall1935representatives}. 
In the bipartite graphs that arise due to our compilation problem, each node has 
the same number of incident edges, a sufficient condition for Hall's matching 
theorem to hold.
Matchings can efficiently be found using the Ford-Fulkerson algorithm 
\cite{Cor2009}.
To this end, one attaches virtual nodes $s$ and $t$ to all nodes on the left and 
right of the bipartite graph, respectively, and determines a minimal network 
flow from $s$ to $t$; see Fig.~\ref{fig:CompProposed}(f).
Finding a minimal flow configuration reveals one matching at a time.
Successively removing edges of a matching from the bipartite graph and 
repeatedly running the Ford-Fulkerson algorithm will reveal all matchings 
$\mathcal{M}_{y}$ with $y=1,...,2m\dy$, as visualized in 
Fig.~\ref{fig:CompProposed}(g).
Finally, to route qubits in step (i), one selects qubits which move to vertical position 
$y$ by iterating over the edges of the $y$th matching $(l,r) \in 
\mathcal{M}_{y}$.
Here, each edge $(l,r)$ signifies that in the generalized column $l$ a qubit 
destined for the generalized column $r$ should be moved to vertical position $y$.
Since edges in each set of matchings $\mathcal{M}_{y}$ point to exactly one final 
destination per vertical position $y$, the fact that $2m\dy$ matchings exist ensures 
that step (i) arranges the qubits in each column so that each row contains 
exactly $m$ qubits with final destination in column $1, 2, 3, ..., 2m\dx$.

\begin{figure}[b]
    \centering
    \includegraphics[width=0.5\textwidth]{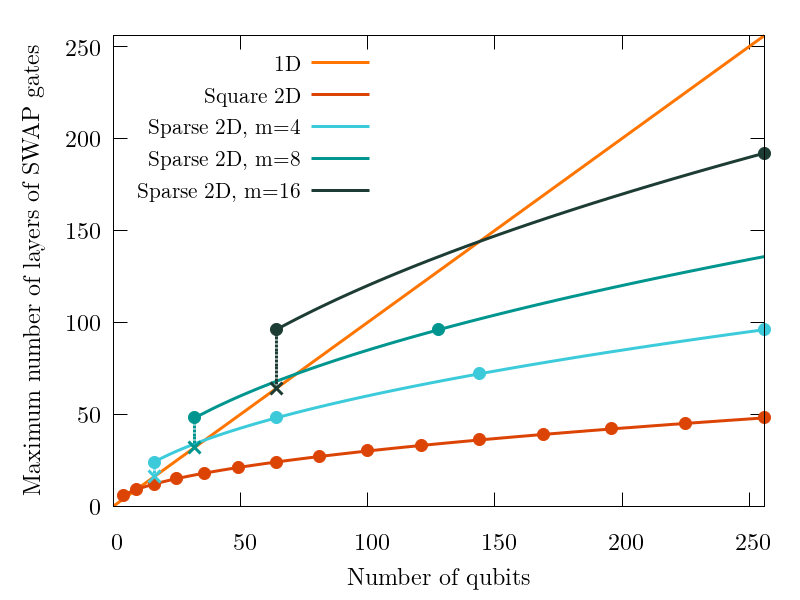}
    \caption{Compilation overhead given by maximum number of layers of 
\textsc{Swap} gates as a function of the number of qubits for (1d) linear qubit 
chains, (2d) rectangular nearest-neighbor grids and the proposed architecture 
for $d_{x}=d_{y}=d$ with $m=4$, 8 and 16.
Lines indicate scaling and dots indicate qubit layout which exist.
Crosses indicate the reduced compilation overhead by using a single parallel 
neighbor sort for devices with $d_{x}=d_{y}=1$. Such devices consist of a square
of qubits, as seen in figure~\ref{fig:sparsedevice}(c).}
    \label{fig:maxswap}
\end{figure}
\paragraph{Compilation overhead}
We close this section by evaluating the compilation overhead.
Considering the succession of parallel neighbor sorts along (i) generalized 
columns of length $2m\dy$, (ii) generalized rows of length $2m\dx$, and (iii) 
again generalized columns of length $2m\dy$, the maximum number of required 
layers of \textsc{Swap} gates is given by
\begin{equation}
  \nd = 4m\dy + 2m\dx
\end{equation}
and the maximum total number of \textsc{Swap} gates by
\begin{equation}
  \ns = 2m\dx\dy(2m\dx + 4m\dy - 3).
\end{equation}
For the specific case $\dx=\dy=d$, and using Eq.~\eqref{eq:NumQubits}, we have
\begin{equation}
    \nd = 6md = 3\sqrt{m\nq}
\end{equation}
and
\begin{equation}
  \ns = 6md^{2}(2md-1) = \frac{3}{2}\nq(\sqrt{m\nq}-1).
\end{equation}
We compare the compilation overhead of the proposed architecture with sparse 2d 
connectivity to the linear and rectangular hardware graphs in 
Fig.~\ref{fig:maxswap}.
Again, the linear configuration (labeled 1D) exhibits a linear scaling of the 
compilation overhead, while the rectangular device with nearest neighbor 
connectivity (labeled 2D) exhibits a square-root scaling.
The compilation overhead of the proposed architecture inherits the favourable 
square-root scaling of 2d hardware topologies, making the proposed architecture 
favorable over 1d devices.
Interestingly, the compilation overhead of the proposed device never exceeds the 
compilation overhead of the linear architecture for $\dx=\dy\geq 2$.
For device structures with $\dx=\dy=1$, the compilation overhead of the 
proposed architecture can be reduced by recognising that the rearrangement of 
the sparse device can always be handled with a single parallel neighbor sort, by 
arranging the qubits along a single line.
This ensures that the compilation overhead for the proposed architecture never 
exceeds the compilation overhead of the 1d architecture even for the smallest 
devices.
This is useful as first experimental realizations of the proposed architecture 
would start from small prototypes.
Finally, for increasing sparsity $m=4, 8, 16, ...$, the compilation overhead of 
the proposed architecture does increase; however, it again never exceeds the overhead 
of the linear device.
This allows for balancing between compilation overhead and 
creating space in between the qubit modules which can be beneficial for 
colocation of classical support electronics.

\section{Discussion}

Inspired by the challenge of scaling up existing silicon quantum hardware, we 
investigate compilation strategies for sparsely-connected 2d qubit arrangements 
and propose a spin-qubit architecture with minimal compilation overhead.
Our considerations are inspired by silicon nanowire split-gate transistors which form finite 1d 
chains of spin-qubits, allowing for the execution of two-qubit operations such 
as \textsc{Swap} gates among neighbors.
Adding to this, we describe a novel silicon junction which can couple up to four
nanowires at one end into 1d or 2d arrangements via spin shuttling and \textsc{Swap} 
operations.
Given these hardware elements, we propose a modular 2d spin-qubit architecture 
with unit cells consisting of diagonally-oriented squares with nanowires along 
the edges and junctions at the corners.

The junction geometry opens space between modules to route the gate lines 
and/or to place cryogenic classical electronics in the quantum processor 
plane~\cite{Vandersypen2017, Boter2019}. Fabricating the qubits and the 
classical control layer using the same technology is appealing because it will 
facilitate the integration process, improving feedback speeds in 
error-correction protocols, and offer potential solutions to wiring and layout 
challenges~\cite{Charbon2016, Franke2019, Xue2021, Park2021, Prabowo2021, 
Ruffino2021}. Integrating classical and quantum devices monolithically, using 
CMOS processes, enables the quantum processor to profit from the most mature 
industrial technology for the fabrication of large-scale 
circuits~\cite{Ruffino2021a}.
We show that this architecture allows for compilation strategies which inherit 
the favourable square-root scaling of compilation overhead in 2d structure 
and outperform the best in class compilation strategy of 1d chains, not only 
asymptotically, but also down to the minimal structure of a single square.
This result shows that scaling silicon nanowires into 2d structures 
will have benefits early on, even in experimental demonstrations of the 
smallest prototypes, thus encouraging building the proposed junction element 
and expanding silicon architectures into 2d arrangements.
%
%
%
%
We further note that our compilation strategies, while being inspired by 
spin-qubits, are equally valid for any other quantum processor with sparse 2d connectivity.

The compilation strategies presented here act to demonstrate the square-root scaling in overhead 
due to routing for the sparsely-connected device, and the advantage of using this device over one 
with a 1d structure. Many other architecture-aware compilation methods exist and show good results
~\cite[e.g.,][]{zulehner2018efficient,childs2019circuit, cowtan2019qubit,li2019tackling,lao2021timing}.
Circuit re-synthesis~\cite{gheorghiu2020reducing} provides another option. 
Alternatively, it may be beneficial to consider a compilation method designed with the desired algorithm 
in mind~\cite[e.g.,][]{Holmes2020,brownecompiler}. However, the distance between qubits will clearly affect 
the overhead introduced through such compilation methods, and thus the results presented here provide 
some indication of their likely performance. Finally, the methods presented in this paper and discussed
above are typically for NISQ-era devices; for the fault-tolerant era, it will clearly be important
to investigate how best to perform error correction on the device.


\subsection{Acknowledgments}
We thank Michael A. Fogarty for his careful read of the manuscript. This research has received funding from the European Union's Horizon 2020 Research and Innovation Programme under grant agreement No 688539 
(http://mos-quito.eu) as well as grant agreement No 951852.
This research has further received funding through the UKRI Innovate UK grant 48482 (NISQ.OS). MFGZ acknowledges support from UKRI Future Leaders Fellowship [grant number MR/V023284/1]. 

\subsection{Competing Interests} The authors declare that they have no 
competing financial interests.

\subsection{Correspondence}
Correspondence and requests for materials should be addressed to O.C. (email: ophelia.crawford@riverlane.com) and M.F.G.Z~(email: mg507@cam.ac.uk).

\begin{widetext}
\section{Appendix A. Two-spin Hamiltonian in a double quantum dot} 

To produce the energy spectrum in Fig.~\ref{Fig:Nanowire}(d), we use the 
following Hamiltonian
\begin{equation}
	H=
	\begin{pmatrix}
		\frac{\varepsilon}{2}+Z_\text{av} & 0 & 0 & 0 & 0\\ 
		0 & \frac{\varepsilon}{2}+\frac{Z_\text{d}}{2} & 0 & 0 & t_\text{s} \\
		0 & 0 & \frac{\varepsilon}{2}-\frac{Z_\text{d}}{2} & 0 & -t_\text{s} \\
		0 & 0 & 0 & \frac{\varepsilon}{2}-Z_\text{av} & 0 \\
		0 & t_\text{s} & -t_\text{s} & 0 & -\frac{\varepsilon}{2}
	\end{pmatrix},
\end{equation}
where $Z_\text{av}$ stands for the Zeeman energy average 
$Z_\text{av}=(g_1+g_2)\mu_\text{B}B/2$ and $Z_\text{d}$ is the Zeeman energy 
difference $Z_\text{d}=(g_1-g_2)\mu_\text{B}B$. Here $g_i$ is the g-factor of 
the particle $i$, $\mu_\text{B}$ is the Bohr magneton and $B$ is the external 
magnetic field. We consider negligible coupling between spins states with different total 
spin number. This Hamiltonian is given in the basis states
$\left\{
\left| \uparrow, \uparrow \right\rangle,
\left| \downarrow, \uparrow \right\rangle,
\left| \uparrow, \downarrow \right\rangle,
\left| \downarrow, \downarrow \right\rangle,
\left| \uparrow\downarrow-\downarrow\uparrow,0 \right\rangle=\left| 
S(20)\right\rangle
\right\}$, assuming $g_1<g_2$.

\section{Appendix B. Timescale of the shuttling sequence}

For silicon electron spin qubits, coherent shuttling has been performed in 8~ns or longer~\cite{Yoneda2021}. However, the lower end of this demonstration has been limited by the control hardware and hence faster coherent shuttling may be achieved. Considering state leakage due to non-adiabatic tunneling as the mechanism for loss of shuttling fidelity at short timescales, we estimate the minimum duration of the shuttling sequence. We consider the probability of a Landau-Zener transition

\begin{equation}
	P_\text{LZ}=\text{exp}\left[-\frac{\pi(2t_\text{s})^2}{2v}\right]
\end{equation}

\noindent where $v$ is the driving velocity across the double QD anticrossing. Considering a minimum pulse amplitude of $4(2t_\text{s})$ and $t_\text{s}\approx 50$~GHz~\cite{Yoneda2021}, we obtain a shuttling time, $t_\text{sh}=235$~ps for $P_\text{LZ}=10^{-4}$. Considering 6 shuttling steps, the total shuttling time corresponds to $T_\text{sh}=1.4$~ns.  

Comparing this figure with the state-of-the-art exchange gate in silicon performed in subnanosecond timescales (~0.8~ns)~\cite{He2019}, there may be regimes in which shuttling time may add a sizable overhead. However, slower exchange gates (controlled by reducing the QD exchange coupling) or synthesized $\sqrt{\textsc{Swap}}$, from single qubit gates and a CPhase gate ($\approx 100$~ns~\cite{xue2021a,noiri2021}) may substantially increase the control fidelity. In that regime, shuttling time becomes negligible. 

We note that the number of shuttling steps could be reduced to 4 by shuttling simultaneously two electrons and performing the SWAP gate when one electron is located under the central gate of the junction and the other under one of the neighboring gates.   

\section{Appendix C. Distance averaging}
In this section, we will calculate the average distance between two qubits in different layouts by summing over all pairs of qubits and dividing by the total number of pairs of qubits. The average distance, $\bar{l}$, is therefore\begin{equation}
    \bar{l} = \frac{2}{n(n-1)} l_{tot},
\end{equation}

where $l_{tot}$ is the total distance between all pairs of qubits. We will make use of the standard sums, which are

\begin{align}
    S_{0} &= \sum_{i=x_{1}}^{x_{2}} 1 = (x_{2} - x_{1} + 1) \label{eq:constant}, \\
    S_{1} &= \sum_{i=x_{1}}^{x_{2}} i = \frac{1}{2}x_{2}(x_{2}+1) - \frac{1}{2}x_{1}(x_{1}-1), \\
    S_{2} &= \sum_{i=x_{1}}^{x_{2}} i^{2} = \frac{1}{6}x_{2}(x_{2}+1)(2x_{2}+1) - \frac{1}{6}x_{1}(x_{1}-1)(2x_{1}-1), \\
    S_{3} &= \sum_{i=x_{1}}^{x_{2}} i^{3} = \frac{1}{4}x_{2}^{2}(x_{2}+1)^{2} - \frac{1}{4}x_{1}^{2}(x_{1}-1)^{2}, \\
    S_{4} &= \sum_{i=x_{1}}^{x_{2}} i^{4} = \frac{1}{30}x_{2}(x_{2}+1)(2x_{2}+1)(3x_{2}^{2}+3x_{2}-1) - \frac{1}{30}x_{1}(x_{1}-1)(2x_{1}-1)(3x_{1}^{2}-3x_{1}-1).
\end{align}

\subsection{Sparsely-connected two-dimensional layout}
\label{app:sparse}
We will find it useful to define the qubits by the particular square they are in, specified by an $x$ co-ordinate with possible values from 1 to $d_{x}$ and a $y$ co-ordinate with possible values from 1 to $d_{y}$. We will use $(a_{x}, a_{y})$ for these co-ordinates, and further use $(b_{x}, b_{y})$ when considering two qubits simultaneously. We will then split the possible pairs of qubits into two -- those with $|a_{x} - b_{x}| \neq |a_{y} - b_{y}|$ and those with $|a_{x} - b_{x}| = |a_{y} - b_{y}|$.

In the calculations that follow, we will assume $d_{x} \geq d_{y}$. A device can be rotated to ensure this is true.

\subsubsection{Pairs of qubits with $|a_{x} - b_{x}| \neq |a_{y} - b_{y}|$}
When considering these pairs of qubits, we split them again into those with $|a_{x} - b_{x}| < |a_{y} - b_{y}|$ and those with $|a_{x} - b_{x}| > |a_{y} - b_{y}|$. We first consider the former. In this case, the shortest distance between any two vertices of the two chosen squares is $|a_{y} - b_{y}|-1$. The shortest path between any one qubit in one square and any other qubit in the other square includes this shortest path between vertices. We will therefore begin by summing over the distances between the nearest vertices of the squares before later including the additional distance to the qubits.

We find that the total sum of distances between pairs of squares is given by
\begin{align}
    l_{sq1} &= \sum_{a_{y}=1}^{\dy} \sum_{b_{y}=a_{y}+1}^{\dy} 2m(b_{y} - a_{y} - 1) \left [ \sum_{i=1}^{b_{y}-a_{y}} 2(\dx - i + 1) - \dx \right ] \nonumber \\
    &= \sum_{f_{y}=1}^{d_{y}} \sum_{c_{y}=1}^{f_{y}-1}2m(c_{y}-1)\left [ \sum_{i=1}^{c_{y}} 2(\dx - i + 1) - \dx \right ] \nonumber \\
    &= \sum_{f_{y}=1}^{d_{y}} \sum_{c_{y}=1}^{f_{y}-1}2m(c_{y}-1)\left [ c_{y}(2\dx + 1 - c_{y}) - \dx \right ] \nonumber \\
    &= m\sum_{f_{y}=1}^{d_{y}} \sum_{c_{y}=1}^{f_{y}-1}\left [ -c_{y}^{3} + c_{y}^{2}(2\dx+2) - c_{y}(3\dx+1)+\dx \right ] \nonumber \\
    &= \sum_{f_{y}=1}^{d_{y}} 2m \left [ -\frac{1}{4}f_{y}^{2}(f_{y}-1)^{2} + \frac{1}{3} f_{y} (f_{y}-1)(2f_{y}-1)(\dx+1) - \frac{1}{2}f_{y}(f_{y}-1)(3\dx+1) + (f_{y}-1)\dx\right ] \nonumber \\
    &= 2m \sum_{f_{y}=1}^{\dy} \left [ -\frac{1}{4}f_{y}^{4} + \frac{1}{6}(7+4\dx)f_{y}^{3} - \frac{1}{4}(7+10\dx)f_{y}^{2} + \frac{1}{6}(5 + 17\dx)f_{y} - \dx\right ] \nonumber \\
    &= -\frac{m}{60}\dy(\dy+1)(2\dy+1)(3\dy^{2}+3\dy-1) + \frac{m}{12}(7+4\dx)\dy^{2}(\dy+1)^{2} - \frac{m}{12}(7+10\dx)\dy(\dy+1)(2\dy+1) \nonumber \\
    &\quad+ \frac{m}{6}(5+17\dx)\dy(\dy+1) - 2\dx\dy \nonumber \\
    &= \frac{1}{30}m \dy (\dy-1)(\dy-2)(4 + (1+10\dx)\dy - 3\dy^{2}).
\end{align}

We now consider qubits with $|a_{x} - b_{x}| > |a_{y} - b_{y}|$. In this case, the shortest distance between any two vertices of the two chosen squares is $|a_{x} - b_{x}|-1$. Again, the shortest path between any one qubit in one square and any other qubit in the other square includes this shortest path between vertices. We will therefore now sum over the distances between the nearest vertices of the squares before later including the additional distance to the qubits. We find this total distance is given by
\begin{equation}
    l_{sq2} = l_{p1}+l_{p2}+l_{p3},
\end{equation}
where
\begin{align}
    l_{p1} &= \sum_{a_{x}=\dx-\dy+1}^{\dx} \sum_{b_{x}=a_{x}+1}^{\dx} 2m(b_{x}-a_{x}-1)\left[\sum_{i=1}^{b_{x}-a_{x}} 2(\dy-i+1)-\dy \right ], \\
    l_{p2} &= \sum_{a_{x}=1}^{\dx-\dy} \sum_{b_{x}=a_{x}+1}^{a_{x}+\dy} 2m(b_{x}-a_{x}-1)\left[\sum_{i=1}^{b_{x}-a_{x}} 2(\dy-i+1)-\dy \right ], \\
    l_{p3} &= \sum_{a_{x}=1}^{\dx-\dy} \sum_{b_{x}=a_{x}+\dy+1}^{\dx} 2m(b_{x}-a_{x}-1)\left[\sum_{i=1}^{\dy} 2(\dy-i+1)-\dy \right ].
\end{align}
Evaluating these in turn, we find
\begin{align}
    l_{p1} &= \sum_{f_{x}=1}^{\dy}\sum_{c_{x}=1}^{f_{x}-1} 2(c_{x}-1)\left [ \sum_{i=1}^{c_{x}} 2m(\dy - i + 1) - \dy \right ] \nonumber \\
    &= \frac{1}{30} m\dy (\dy-1)(\dy-2)(4 + (1+10\dy)\dy - 3\dy^{2}) \nonumber \\
    &= \frac{1}{30} m\dy (\dy-1)(\dy-2)(7\dy^{2}+\dy+4),
\end{align}
\begin{align}
    l_{p2} &= 2\sum_{a_{x}=1}^{\dx-\dy} \sum_{c_{x}=1}^{\dy}m(c_{x}-1) \left [ \sum_{i=1}^{c_{x}}2(\dy-i+1)-\dy \right ] \nonumber \\
    &= \sum_{a_{x}=1}^{\dx-\dy} 2m \left [ -\frac{1}{4}\dy^{2}(\dy+1)^{2} + \frac{1}{3} \dy (\dy+1)^{2}(2\dy+1) - \frac{1}{2}\dy(\dy+1)(3\dy+1) + \dx\dy\right ] \nonumber \\
    &=  2m\dy(\dx-\dy) \left [ -\frac{1}{4}\dy(\dy+1)^{2} + \frac{1}{3} (\dy+1)^{2}(2\dy+1) - \frac{1}{2}(\dy+1)(3\dy+1) + \dx\right ] \nonumber \\
    &= \frac{1}{6}m\dy(\dy-1)(\dx-\dy)(5\dy^{2}+\dy+2),
\end{align}
\begin{align}
    l_{p3} &= 2m\dy^{2} \sum_{a_{x}=1}^{\dx-\dy} \sum_{c_{x}=1}^{\dx-\dy-a_{x}}(c_{x}+\dy-1) \nonumber \\
    &= 2m\dy^{2} \sum_{a_{x}=1}^{\dx-\dy}\left [ \frac{1}{2}(\dx-\dy-a_{x})(\dx-\dy-a_{x}+1)+(\dx-\dy-a_{x})(\dy-1)\right ] \nonumber \\
    &= m\dy^{2}\sum_{a_{x}=1}^{\dx-\dy} [a_{x}^{2} + (2\dx-1)a_{x}+(\dx-\dy)(\dx+\dy-1)] \nonumber \\
    &= m\dy^{2} \left [\frac{1}{6}(\dx-\dy)(\dx-\dy+1)(2\dx-2\dy+1) - \frac{1}{2}(2\dx-1)(\dx-\dy)(\dx-\dy+1) +(\dx-\dy)^{2}(\dx+\dy-1) \right] \nonumber \\
    &= \frac{1}{3}m\dy^{2}(\dx-\dy)(\dx-\dy-1)(\dx+2\dy-2).
\end{align}
We now have the total distance between all squares for which $|a_{x}-b_{x}| \neq |a_{y}-b_{y}|$. As each square contains $4m$ qubits, we first need to multiply these distances between the squares by $16m^{2}$. We also need to add the additional distance due to the fact the qubits are some distance from the vertex involved in the shortest distance between vertices. For each pair of squares, we therefore need to add the distance
\begin{equation}
    l_{q} = 4 \sum_{i=1}^{2m} \sum_{j=1}^{2m} (i+j-1) = 4m \sum_{i=1}^{2m} [2i + 2m -1] = 32m^{3}.
\end{equation}
The total number of pairs of squares is
\begin{equation}
    n_{sq} = \frac{1}{2}\dx\dy(\dx\dy-1).
\end{equation}
The total number of pairs of squares with $|a_{x} - a_{y}| = |b_{x}-b_{y}|$ is
\begin{align}
    n_{sq1} &= 4\sum_{i=1}^{\dy-1} \sum_{a_{y}=1}^{i} \sum_{b_{y}=a_{y}+1}^{i} 1 + 2(\dx-\dy+1)\sum_{a_{y}=1}^{\dy} \sum_{b_{y}=1}^{\dy} 1 \\
    &= 2\sum_{i=1}^{\dy-1} (i^{2}-i) + \dy(\dy-1)(\dx-\dy+1) \\
    &= \frac{1}{3}\dy(\dy-1)(2\dy-1) - \dy(\dy-1) + \dy(\dy-1)(\dx-\dy+1) \\
    &= \frac{1}{3}\dy(\dy-1)(3\dx-\dy-1).
\end{align}

Therefore, the total number of pairs of squares with $|a_{x} - a_{y}| \neq |b_{x}-b_{y}|$ is
\begin{align}
    n_{sq2} = n_{sq} - n_{sq1} = \frac{1}{6}\dy[3\dx^{2}\dy - 6\dx\dy+3\dx+2\dy^{2}-2].
\end{align}
The total distance between qubits for which $|a_{x} - a_{y}| \neq |b_{x}-b_{y}|$ is finally given by
\begin{equation}
    d_{neq} = 16m^{2}(l_{sq1}+l_{sq2}) + 32m^{3}n_{sq2}.
\end{equation}

\subsubsection{Pairs of qubits with $|a_{x} - b_{x}| = |a_{y} - b_{y}|$}
We now consider pairs of qubits with $|a_{x} - b_{x}| = |a_{y} - b_{y}|$. In this case, there is not a unique pair of nearest vertices in the two squares and so we must consider different pairs of qubits in different manners.

We will consider squares with a particular value of $(b_{x} - a_{x})$ in turn, giving us a line of diagonally oriented squares. For the moment, we assume that we have a set of $n$ such squares; we will sum over the different values of $n$ later. We first consider summing over the distances between all pairs of qubits down the sides of the squares, given by
\begin{align}
    l^{n}_{ss} &= 2 \sum_{i=1}^{n} \sum_{j=i+1}^{n} \sum_{k=1}^{m} \sum_{l=1}^{m} \left [ 2m(j-i) + (l-k) \right ] \nonumber \\
    &= 4m^{3} \sum_{i=1}^{n} \sum_{j=i+1}^{n} (j-i) \nonumber \\
    &= 4m^{3} \sum_{i=1}^{n} \left [ \frac{1}{2}n(n+1) - \frac{1}{2}i(i+1) -in + i^{2}\right ] \nonumber \\
    &= 4m^{3} \sum_{i=1}^{n} \left [ \frac{i^{2}}{2} - \frac{i}{2}(2n + 1) + \frac{1}{2}n(n+1) \right ] \nonumber \\
    &= 4m^{3} \left [ \frac{1}{12}n(n+1)(2n+1) - \frac{1}{4}n(n+1)(2n+1) + \frac{1}{2}n^{2}(n+1) \right ] \nonumber \\
    &= \frac{2}{3}m^{3}n(n-1)(n+1).
\end{align}

We now consider pairs of qubits on opposite sides of the line of squares; however, to avoid double counting, we do not include pairs of qubits on opposite sides of the same square. This distance sum is given by
\begin{align}
    l^{n}_{os} &= 2\sum_{i=1}^{n} \sum_{j=i+1}^{n} \sum_{k=1}^{m} \sum_{l=1}^{m} \left [ 2m(j-i) + (l-k) + m \right ] \\
    &= l^{n}_{ss} + m^{3}n(n-1) \\
    &= \frac{1}{3}m^{3}n(n-1)(2n+5).
\end{align}
We next consider pairs of qubits where one of the pair is on one of the sides of the line of squares and the other is on an edge at right angles to this, on a  `rung' of the line of squares. Again, to avoid double counting, we exclude some pairs of qubits here. This sum of distances is given by
\begin{align}
    l^{n}_{sr} &= 2\sum_{i=1}^{n} \sum_{j=1}^{2i-2} \sum_{k=1}^{m}\sum_{l=1}^{m} \left [(2i-j-1)m+l+k-1 \right] + 2\sum_{i=1}^{n} \sum_{j=2i}^{2n}\sum_{k=1}^{m}\sum_{l=1}^{m} \left [ (j-2i+1)m + l-k\right ] \nonumber \\
    &= 2\sum_{i=1}^{n}\sum_{j=1}^{2i-2} \left [ (2i-j-1)m^{3} + m^{2}(m+1) - m^{2} \right ]+ 2m^{3}\sum_{i=1}^{n}\sum_{j=2i}^{2n} (j-2i+1) \nonumber \\
    &= 2m^{3} \sum_{i=1}^{n} \sum_{j=1}^{2i-2} (2i-j)+ 2m^{3}\sum_{i=1}^{n} \sum_{k=1}^{2n-2i+1} k \nonumber \\
    &= 2m^{3} \sum_{i=1}^{n} \left [ 2i(2i-2) - \frac{1}{2}(2i-2)(2i-1) + (2n-2i+1)(2n-2i+2) \right ] \nonumber \\
    &= m^{3} \sum_{i=1}^{n} \left [ 8i^{2} - 8i(n+1) + 2n^{2}+3n \right ] \nonumber \\
    &= m^{3} \left [ \frac{4}{3}n(n+1)(2n+1) - 4n(n+1)^{2} +2n^{3} + 3n^{2}\right ] \nonumber \\
    &= \frac{2}{3} m^{3}n (4n^{2} + 3n - 4). 
\end{align}
We finally consider pairs of qubits where both qubits are on a `rung' of the line of squares. Here, the total distance between pairs of qubits is given by
\begin{align}
    l^{n}_{rr} &= 2n \sum_{k=1}^{m}\sum_{l=k+1}^{m} (l-k) + \sum_{i=1}^{2n} \sum_{j=i+1}^{2n} \sum_{k=1}^{m} \left [ \sum_{l=1}^{m-k+1} [(j-i)m+k+l-1] + \sum_{l=m-k+2}^{m} [(j-i+2)m -k-l+1] \right ] \nonumber \\
    &= 2n \sum_{k=1}^{m} \sum_{l=k+1}^{m} (l-k) + \sum_{i=1}^{2n}\sum_{j=i+1}^{2n} \sum_{k=1}^{m} \left [ \sum_{l=1}^{m} (j-i)m + \sum_{l=1}^{m-k+1}(k+l-1) + \sum_{l=m-k+2}^{m} (2m - k - l+1) \right ] \nonumber \\
    &= \frac{1}{3}nm(m+1)(m-1) + \frac{1}{6}m^{3}2n(2n+1)(2n-1) \nonumber \\
    & \quad + \sum_{i=1}^{2n}\sum_{j=i+1}^{2n} \sum_{k=1}^{m} \left [\frac{1}{2}(m-k+1)(m+k) + (2m-k+1)(k-1) - \frac{1}{2}m(m+1) + \frac{1}{2}(m-k+1)(m-k+2) \right ] \nonumber \\
    &= \frac{1}{3}nm(m+1)(m-1) + \frac{1}{3}m^{3}n(2n+1)(2n-1) + \sum_{i=1}^{2n}\sum_{j=i+1}^{2n} \sum_{k=1}^{m} \left [ -k^{2}+(m+1)k + \frac{1}{2}m(m-1) \right ] \nonumber \\
    &= \frac{1}{3}nm(m+1)(m-1) + \frac{1}{3}m^{3}n(2n+1)(2n-1) + n(2n-1)\left [ -\frac{1}{6}m(m+1)(2m+1) + \frac{1}{2}m(m+1)^{2} + \frac{1}{2}m^{2}(m-1)\right ] \nonumber \\
    &= \frac{2}{3}nm(2nm^{2}+2n^{2}m^{2}-m^{2}+n-1).
\end{align}

Taking the sum $l^{n} = l^{n}_{ss} + l^{n}_{os}+l^{n}_{sr}+l^{n}_{rr}$, we find
\begin{equation}
    l^{n} = \frac{1}{3}nm(13nm^{2}+16n^{2}m^{2}-17m^{2}+2n-2).
\end{equation}
We now need to sum over the different values of $n$. There are four lines of squares for each $1 \leq n \leq d_{y} - 1$ and $2(\dx-\dy+1)$ squares with $n=\dy$. Therefore, the total distance between qubits with $|a_{x}-b_{x}| = |a_{y}-b_{y}|$ is
\begin{align}
    d_{eq} &= \frac{4}{3}m \sum_{n=1}^{\dy-1} (13n^{2}m^{2} + 16n^{3}m^{2}-17nm^{2}+2n^{2}-2n) + \frac{2}{3}(\dx-\dy+1)m\dy(13m^{2}\dy+16m^{2}\dy^{2}-17m^{2}+2\dy-2) \nonumber \\
    &= \frac{4}{3}m\dy(\dy-1) \left [ \frac{1}{6}(13m^{2}+2)(2\dy-1) + 4 m^{2}\dy(\dy-1) - \frac{1}{2}(17m^{2}+2)\right ] \nonumber \\
    & \quad + \frac{2}{3}m\dy(\dx-\dy+1)(13m^{2}\dy+16m^{2}\dy^{2}-17m^{2}+2\dy-2).
\end{align}

\subsubsection{Mean distance}
The mean distance between two randomly selected qubits is given by
\begin{align}
    \bar{l} &= \frac{2}{\nq(\nq-1)}(d_{neq}+d_{eq}), \label{eq:lbarsum}
\end{align}
where we recall
\begin{equation}
    \nq = 4m\dx\dy.
\end{equation}

\subsection{Linear layout}
\label{app:1d}
Given $m$ qubits laid out linearly, the average distance between a pair of qubits is
\begin{align}
    \bar{l} &= \frac{2}{m(m-1)} \sum_{i=1}^{m}\sum_{j=i+1}^{m} (j-i) \\
    &= \frac{1}{3m(m-1)} m(m+1)(m-1) \\
    &= \frac{1}{3}(m+1).
\end{align}

\subsection{Rectangular two-dimensional layout}
\label{app:full}
Given a $\lx \times \ly$ rectangle of qubits, the average distance between a pair of qubits is given by
\begin{align}
    \bar{l} &= \frac{2}{\lx\ly(\lx\ly-1)} \sum_{i=1}^{\lx} \sum_{j=1}^{\ly} \left [ \sum_{k=i+1}^{\lx} \sum_{l=j}^{\ly} (k+l-i-j) + \sum_{k=1}^{i} \sum_{l=j+1}^{\ly} (i + l - k - j) \right ]. \label{eq:2dsum}
\end{align}
$(i,j)$ and $(k,l)$ are the coordinates of two points. We sum over all possible $i$ and $j$, and, to avoid double counting, sum over the distance between the point $(i,j)$ and any points on the same row and to the right, or on a higher row. The first term inside the brackets sums over those points to the right and on the same row or a higher row, whilst the second term sums over those points in the same column or to the left on a higher row. Rearranging terms, we find
\begin{align}
    l_{tot} &= \sum_{i=1}^{\lx} \sum_{k=1}^{\lx} \sum_{j=1}^{\ly} \sum_{l=j+1}^{\ly} (l-j) + \sum_{j=1}^{\ly} \sum_{l=j}^{\ly} \sum_{i=1}^{\lx} \sum_{k=i+1}^{\lx} (k-i) + \sum_{j=1}^{\ly} \sum_{l=j+1}^{\ly} \sum_{i=1}^{\lx} \sum_{k=1}^{i} (i-k) \nonumber \\
    &= \frac{1}{6}\lx^{2} \ly(\ly-1)(\ly+1) + \frac{1}{12}\lx\ly(\lx-1)(\lx+1)(\ly+1) + \frac{1}{12}\lx\ly(\lx-1)(\ly-1)(\lx+1) \nonumber \\
    &= \frac{1}{12}\lx\ly[2\lx(\ly^{2}-1) + (\lx^{2}-1)(\ly+1) + (\lx^{2}-1)(\ly-1)] \nonumber \\
    &= \frac{1}{6}\lx\ly(\lx\ly - 1)(\lx+\ly),
\end{align}
and so
\begin{equation}
\bar{l} = \frac{1}{3}(\lx+\ly).
\end{equation}
\end{widetext}


\bibliographystyle{naturemag}


\end{document}